Communication

# What makes an explosion happen?

Chang Q Sun[1,2,]*, Chuang Yao[1], Lei Zhang[3,4], Yongli Huang[5]

**Highlight**

- The lone pair ":" and dangling $H^+$ are the primary functional unit for molecular interaction
- Excessive $H^+$ or lone pairs derive $H \leftrightarrow H$ anti-HB or $X{:}\Leftrightarrow{:}Y$ super-HB molecular repulsion
- X:H-Y tension and intermolecular repulsion govern the structure stability and energy storage
- The absence/presence of the X:H-Y tension fosters the spontaneous/constrained detonation

**Abstract**

Energy density and structure stability are of key concerns in devising energy materials such as CNHO molecular crystals and the emerging *cyclo*-$N_5^-$ complexes for desired explosive functionalities with a mechanism being open for exploration. An extension of the recent progress in solvation suggests that a combination of the intermolecular hydrogen bond (X:H–Y or HB with ':' being electron lone pair of X, Y = O or N ) tension and the super-HB ($X{:}\Leftrightarrow{:}Y$) or anti-HB ($H\leftrightarrow H$) compression not only stabilizes the structure but also stores excessive energy by shortening the intramolecular bonds. The presence of the X:H constrains and the presence of the $X{:}\Leftrightarrow{:}Y$ and/or $H\leftrightarrow H$ fosters the explosion. Observations suggest that the absence of X:H–Y tension results in the spontaneous aquatic explosion of alkali metals and molten alkali halides. The lack of $X{:}\Leftrightarrow{:}Y$ repulsion fosters no explosion of the molten NaCl in liquid $NH_3$, nor the molten $Na_2CO_3$ salt or $H_3BO_3$ acid in water. The findings shall offer guidelines for devising efficient energy materials and reconciling the nature origin of water ice, aqueous solutions, and explosive energy materials – significance of electron lone pairs and protons.

[1] EBEAM, School of Materials Science and Engineering, Yangtze Normal University, Chongqing 408100, China (Yaoc@yznu.cn)
[2] NOVITAS, School of Electrical and Electronic Engineering, Nanyang Technological University, 639798, Singapore; (ecqsun@ntu.edu.sg; *corresponding author)
[3] Software Center for High Performance Numerical Simulation, Institute of Applied Physics and Computational Mathematics, Beijing, 100088, China (zhang_lei@iapcm.ac.cn)
[4] Laboratory of Computational Physics, Institute of Applied Physics and Computational Mathematics, Beijing, 100088, China
[5] School of Materials Science and Engineering, Xiangtan University, Xiangtan 411105 China (Huangyongli@xtu.edu.cn)

**Content entry**

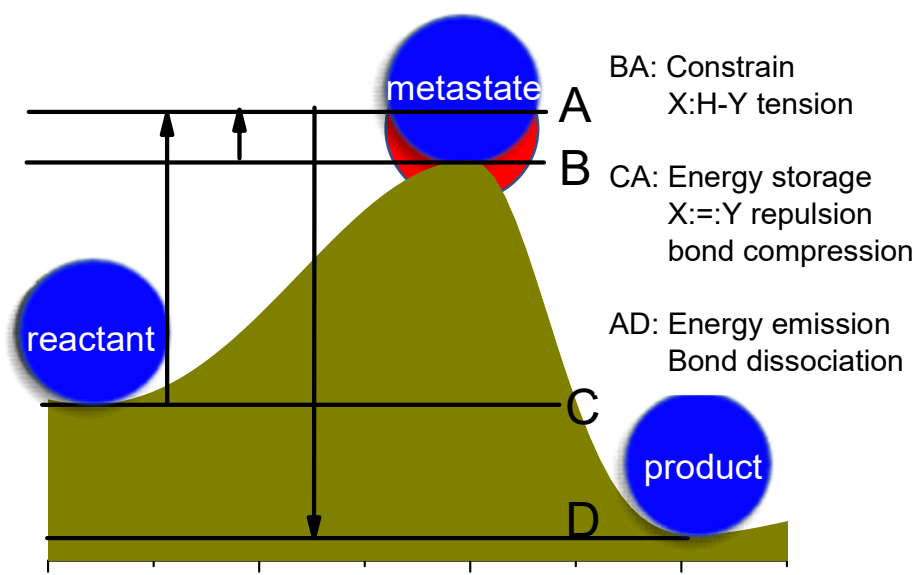

**The combination of the hydrogen-bond (X:H–Y) tension and the super-HB (X:⇔:Y) or the anti-HB (H↔H) repulsion not only stabilizes the structure but also stores energy by shortening the intramolecular covalent bonds. The X:H tension constrains and the X:⇔:Y repulsion fosters explosion.**

Explosions can be categorized into two types – constrained and spontaneous. The traditional CNHO molecular crystals such as TNT and TATB [1] and the emerging *cyclo*-$N_5^-$:($4H_3O^+$; $2H_3O^+ + 2NH_4^+$) complexes [2-5] undergo constrained explosion. One often expects high energy density and low sensitivity to a perturbation such as thermal excitation or shock wave impulse for practical applications. Once ignited, detonation takes place and the carrier erupts energy vigorously, impacting as heavy as possible to the targeted object. The manner, velocity, and amount of energy eruption is subject to its initial storage and the structural stability of the carrier molecular crystals, which can be controlled by adjusting the relative number of protons and lone pairs carried by a molecular motif [1].

Contrastingly, alkali metals (Z = Li, Na, K, Rb, Cs)[6] and molten alkali halides (ZΓ; Γ = F, CL, Br, I) such as NaCl and LiBr [7, 8] explode spontaneously when contacting to water – called aquatic spontaneous explosion. When reacting with water at room temperature, metallic Na and molten NaCl explode rigorously with production of $H_2$, NaOH, shock waves and heat that ignites the $H_2$ [6, 8, 9]. In fact, the bulk Na solvation proceeded in four stages in a few seconds [8]. Firstly, it dissolves gradually to dye the water into purple color; secondly, the Na releases gas/smoke with flame; and followed by abrupt explosion. Dropping a drop of liquid NaK alloy into liquid water at the ambient shoots spikes explosively out 0.2-0.4 ms immediately after the droplet contacting water [6, 7], see photos in Figure 1a. This explosion process is too fast to be attributed to the heat repellant, instead, "alkali metal atoms at the droplet surface each loses its valence electron within picoseconds. These electrons dissolve in water and react in pairs to form molecular $H_2$ and $OH^-$ hydroxide" [6]. Likewise, solvation of or pouring the molten NaCl ($T_m$ = 801 °C) on ice causes massive explosion and leaves a cave behind the ice. The aquatic explosion of the molten NaCl readily breaks a fish tank holding water inside. Figure 1 shows photos captured from the aquatic explosion of the NaK droplets in water and the molten NaCl poured in water and on ice [6, 7].

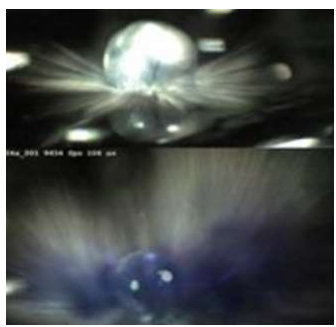 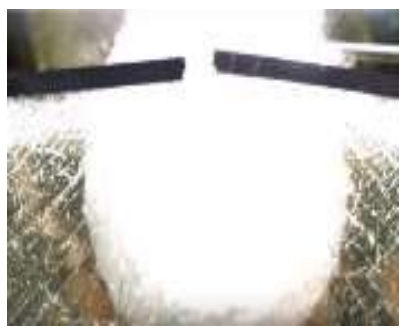 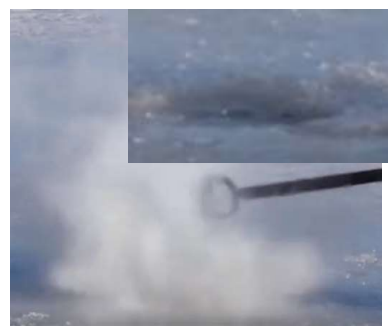

(a)　　　　　　　　　　　　　(b)　　　　　　　　　　　　　(c)

Figure 1. Unconstrained aquatic explosion of (a) liquid NaK alloy droplet when contacting water [6], molten NaCl poured into (b) a water-filled fish tank and (c) on ice. Explosions create spikes abruptly and the molten NaCl explosion breaks the fish tank [7] and (b) leaves a cave behind ice (Tencent video, public domain).

However, replacing the NaK alloy with a drop of molten Al at ~1,000 °C dropping in water, or pouring molten Cu at 1083°C on ice, produces neither spark nor explosion, instead, displaying the Leidenfrost effect [10] and Rayleigh instability [11]. The molten Al or Cu just melts and spreads on ice. The steam separates the liquid metals and water surface on a 0.1 sec timescale. In contrast, solvation of molten $Na_2CO_3$ ($T_m$ = 851 °C) or molten $H_3BO_3$ ($T_m$ = 171 °C) does not foster any explosion though the melting temperatures of these two compounds are different from that of molten NaCl though the reactions also produce $Na^+$ and $H^+$ ions. Replacing $H_2O$ with −77.8 °C liquid $NH_3$ could form less-apparent transient spikes but the liquid ammonia do not foster any explosion [6].

Much focus has been given on the energy dissipation dynamics of the detonation, but limited attention is paid to the understanding of structure stability and energy storage. An examination of the molten NaCl explosive fragmentation in water suggested that the lower viscosity of the liquid salts foresters the explosion [12]. Density functional theory computations [13] suggested that the nonlocal X:H van der Waals dispersive interaction is dominant to the cohesive energies of the molecular crystals without the intramolecular or their coupling being involved. The intermolecular weak X:H dispersion force may play an important role in stabilizing the structures [14]. By analyzing a large volume of experimental and theoretical studies, Tsyshevsky and co-workers [15] recommended that: (i) the chemical composition and structure configuration define the chemistry and stability of the crystal; (ii) the form of and strength of intermolecular interactions are critical; (iii) the morphological defects are imperative for reliable predictions of structural stability. Unfortunately, it is unclear what discriminates the constrained from the spontaneous aquatic explosions. Questions also arise what stabilizes the molecular structure and how to store energy into the molecular crystals at the atomic and molecular scales. What is the reaction dynamics of the aquatic spontaneous explosion in addition to the initiation by the cationic Coulomb fission [6]. For the X:H−Y bonding systems such as water, the X:H nonbond dissipates energy caped at the $10^{-1}$ eV level while the intramolecular covalent bond stores and emits energy upon relaxation by solvation [16] or heating cycling [17].

Recent progress in water ice [18, 19] and solvation dynamics [16, 20] advanced the O:H–O coupled oscillator pair, the repulsive H↔H anti-HB in the acid and alcohol solutions and the compressive O:⇔:O super-HB in the basic and $H_2O_2$ solutions . These entities can be extended to inter- and intramolecular interactions for the explosive organic molecular substance and reaction dynamics by replacing O with X, Y = N or O. As a matter of fact, lone pairs and dangling protons formed the primary structural and functional element to molecular crystals such as foods, drugs, explosives, and even DNA, proteins and cells.

The following describes the principle of X:H-Y relaxation [21], H↔H and X:⇔:Y repulsion [22] and the recommended mechanism for explosion. Because of the Coulomb coupling between electron pairs on the adjacent X and Y of the segmental X:H–Y bond, tension lengthens its X:H nonbond but shortens its H–Y bond [23]. The H↔H and X:⇔:Y repulsion compresses their adjacent intramolecular covalent bonds of the molecular motifs. One can imagine what will happen to the intramolecular covalent bonds if paired up the intermolecular H↔H or X:⇔:Y compression and X:H–Y tension. This combination shortens all the covalent bonds to store energy whose amount depends on the extent of the covalent bond contraction. The combination determines the structural stability and the sensitivity to ignition. Once the tensile X:H breaks, the crystal crashes and the compressed covalent bonds break catastrophically - detonation takes place. The energy release is expected to be coordinated along a single event, as one can see an explosion. Nevertheless, one should be focused on the structure stability and energy storage before calculating the impact or amount of energy eruption at explosion.

An extension of the recent findings on solvation charge injection of Lewis basic solutions [16] and the computations on *clyco*-$N_5^-$:($4H_3O^+$; $3H_3O^+ + 2NH_4^+$) complexes [2] leads to a formulation of the explosions on the factors dictating the stability and energy storage. Applying the X:H–Y tension and H↔H or X:⇔:Y repulsion to the inter- and intra-molecular cooperative interactions of the following species (Z = alkali metal; Γ = halide) and processes, shown in Table 1, clarifies factors controlling the constrained and unconstrained explosions.

Table 1     Criteria for the spontaneous and constrained explosion and non-explosion. The second stage of spontaneous aquatic reaction iterates exactly the ZOH base aqueous solvation dynamics [16]. *

| category | reactant | ⇒ | Reaction Dynamics | Z↔Z | H↔H | X:⇔:Y | X:H–Y |
|---|---|---|---|---|---|---|---|
| spontaneous aquatic explosion | $2Z + nH_2O$ (ambient T) | ⇒ | $2ZOH + heat + (n-2)H_2O + H_2\uparrow$ [$Z^+\leftrightarrow Z^+ + 2(H_2O:\Leftrightarrow:OH^-)$] + heat + $(n-4)H_2O + H_2\uparrow$ | √ | × | √ | × |
| | (molten) $2Z\Gamma + nH_2O$ | ⇒ | $2ZOH + (n-2)H_2O + heat + 2H\Gamma\uparrow$ [$Z^+\leftrightarrow Z^+ + 2(H_2O:\Leftrightarrow:OH^-)$] + heat + $(n-4)H_2O + 2H\Gamma\uparrow$ | √ | × | √ | × |
| Constrained explosion | CHNO | ⇒ | $[n(X:\Leftrightarrow:Y) + m(X:H–Y)] + ...$ | × | × | √ | √ |
| | $cyclo$-$N_5^-$ complexes | ⇒ | $[p(H^+\leftrightarrow H^+) + q(X:H–Y)] + ...$ | × | √ | √ | √ |
| non-explosion | (molten) $2Z + nNH_3$ (-78 °C) | ⇒ | $[Z^+\leftrightarrow Z^+ + H_2N^-:H–NH] + (n-2)NH_3 + H_2\uparrow$ | √ | × | × | √ |
| | (molten) $Na_2CO_3 + nH_2O$ | ⇒ | $[Z^+\leftrightarrow Z^+] + [CO_3]^{2-}\cdot nH_2O$ | √ | × | × | × |
| | (molten) $2H_3BO_3 + nH_2O$ | ⇒ | $3(H^+\leftrightarrow H^+) + 2[BO_3]^{3-}\cdot nH_2O$ | × | √ | × | × |

*m, n, p, q are numbers of the specified intermolecular nonbonds.

Two steps are involved in the aquatic spontaneous explosion. Firstly, $H_2O$ dipoles hydrate the Z alkali metals and the molten $Z\Gamma$ salts as well to dissolve them into constituent ions, $Z^+$ and $\Gamma^-$; the $Z^+$ replaces a $H^+$ of the $H_2O$ to form the basic ZOH and $H_2\uparrow$ gas for the Z metal, and $H\Gamma\uparrow$ for the $Z\Gamma$ salt. Secondly, hydration of the ZOH proceeds and turns the ZOH into the well-defined $Z^+$ and $OH^-$ hydroxide. In basic solvation, each of the $OH^-$ interacts with its neighboring $H_2O$ to form an $OH^-\cdot 4H_2O$ motif in which one O:H–O bond transits into a highly-repulsive O:⇔:O super-HB. This second process repeats exactly what happened upon base solvation [16].

The O:⇔:O compression is much stronger than the critical pressure 1.35 GPa for liquid water transiting into ice at the ambient temperature [24]. The O:⇔:O compression softens the H–O bond in the continuous network of liquid water from 3200 to a dispersed band of $2000 \pm 200$ cm$^{-1}$ and the 1.35 GPa dispersed the same band into $2900 \pm 100$ cm$^{-1}$. One can imagine the power of the O:⇔:O repulsion in the second step of Z and molten $Z\Gamma$ solvation that repeats the same process of basic ZOH hydration [16]. Therefore, solvation of neutral Z metals and the molten $Z\Gamma$ salts creates excessive O:⇔:O repulsion at the expense of an O:H–O bond between the solute and the solvent. Thus, the production of [$Z^+\leftrightarrow Z^+ + 2(H_2O:\Leftrightarrow:OH^-) + ...$] in the basic solution reconciles the unconstrained aquatic explosion without presence of the intermolecular O:H–O tension to equilibrate the O:⇔:O repulsion.

However, the absence of super-HB compression fosters no explosion of the Z metal when contacting liquid $NH_3$. The replacement of an H with a Z transits the $NH_3$ into H + $NH_2$Z and the solvation turns the $NH_2$Z into the $Z^+$ and the $(NH_2)^-$ which has two pairs of lone pairs and two protons. The $(NH_2)^-$ performs the same as an$H_2O$ and then to form a $(NH_2)^-\cdot(4NH_3; 3NH_3 + NH_2^-)$ motifs with its four neighbors. Regularly, each $NH_3$ molecule has three protons and one lone pair. The $NH_3$ interacts with its four-coordinated neighbors through a pair of N:H–N bonds and a pair of H↔H anti-HBs, which explains why ammonia has lower melting temperature at -78 °C. The $(NH_2)^-\cdot 4NH_3$ motif formation up on Z metal solvation turns at least one repulsive H↔H into the N:H–N bond, which stabilizes the ammonia, instead. The transition fosters no explosion but stabilizes the solvent. In contrast, water dissolves the molten $Z_2CO_3$ salt and $H_3BO_3$ acid into the respective $2Z^+ + (HCO_3)^-$ and $H^+ + (H_3BO_3)^-$, being regular cases of complex salt and acid solvation. The ions polarize their neighboring $H_2O$ molecules to form each a hydration volume without super-HB or anti-HB being formed [20], though the $Z^+ \leftrightarrow Z^+$ might be presented. The $H^+$ forms a $H_3O^+$ that turns one O:H–O bond into the repulsive H↔H, being insufficiently strong to foster a repulsion.

Figure 2 exemplifies molecular motifs for (a) the TNT ($2C_7H_5N_3O_6$), (b) the *cyclo*-$N_5^-$:$4H_3O^+$, and (c) the *cyclo*-$N_5^-$: ($3H_3O^+ + 2NH_4^+$) complexes optimized from intensive quantum calculations [2, 3]. The *cyclo*-$N_5^-$ can only be stabilized in acidic conditions of the excessive $(H_3O)^+$ or $(NH_4)^+$ [4, 25-31]. For the TNT in Figure 2a, there are ten protons $H^+$ and $30 = 2\times(2\times 6+3)$ electron lone pairs ":" surrounding the motif. According to the nonbond counting rules [20], protons and lone pairs can interact with their same kinds of its neighboring motifs to form the X:H–Y, X:⇔:Y, or H↔H. There will be 20 HBs and 20 super–HBs per TNT motif with its neighbors. The X:H–Y tension and the X:⇔:Y compression combination stabilizes the crystals and stores energy into the covalent bonds, which ensures the constrained explosion of the TNT and other CNHO energetic crystals, as well.

Figure 2b illustrates the force diagram for the *cyclo*-$N_5^-$:$4H_3O^+$ stabilization and energy storage [2]. The N atom in the *cyclo*-$N_5^-$ ring undergoes the $sp^2$ orbital hybridization with each creating one pair of electron lone pair and one unpaired π electron. These electrons form the attractive and repulsive aromatic components along the *cyclo*-$N_5^-$ ring [2]. At the critical concentration of $4H_3O^+$ surrounding the *cyclo*-$N_5^-$, the circumferential H↔H repulsion stretches the radial N:H–O bonds between the $H_3O^+$ and the $N_5^-$ ions. The stretching N:H–O shortens its H–O parts and lengthens the N:H part, because of the unique feature of the hydrogen-bond cooperative relax-ability. The N:H stretching weakens the compressive aromatic component and reduces the inter-lone-pair repulsion along the $N_5^-$ ring, which shortens the N–N bond to gain energy. As shown in Figure 2c for the bond relaxation of the *cyclo-*$N_5^-$:($3H_3O^+ + 2H_4N^+$) complex [3], the N–N bond contracts to 1.32 Å indeed from 1.38 for the *cyclo-*$N_5^-$ without acidic entrapment. At the same time, the N:H expands to 2.10 Å for the N:H–N bond and to values in the range of 2.17-2.26 Å for the N:H–O bonds, [2, 3] in contrast to the standard O:H length of 1.70 Å at 4 °C.

Hence, both the circumferential $H^+\leftrightarrow H^+$ repulsion and the radial N:H–O or the N:H–N tension shorten all the H–O/N and the N–N bonds, which not only stabilizes the *cyclo-*$N_5^-$ complexes but also stores excessive energy to raise the reactant to the energetic metastable state, as illustrated in Figure 3. This configuration clarifies the reason why the acidic entrapment could stabilize the *cyclo-*$N_5^-$ [25-29] and why the X:H–Y and X:⇔:Y or H↔H coexistence is essential to the constrained explosion. The X:H breaking initiates and the X:⇔:X fosters the explosion of the *cyclo-*$N_5^-$ complexes. In the *cyclo-*$N_5^-$ complexes, the H↔H repulsion plays a part of stabilization but fosters no explosion as the H↔H does in the molten $H_3BO_3$ acid solvation.

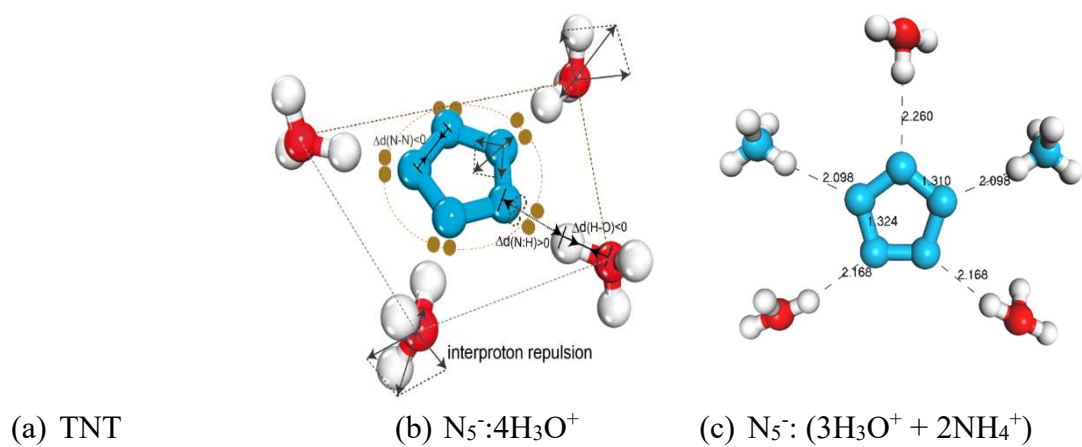

(a) TNT  (b) $N_5^-$:$4H_3O^+$  (c) $N_5^-$: ($3H_3O^+ + 2NH_4^+$)

Figure 2 Molecular motifs for (a) the TNT ($2C_7H_5N_3O_6$), (b) the *cyclo-*$N_5^-$:$4H_3O^+$ with the force diagram, and (c) the *cyclo-*$N_5^-$:($3H_3O^+ + 2NH_4^+$) with bond relaxation. For TNT, the black spheres are carbon, small grey spheres are H, red ones are oxygen and the blue ones are N. Each N carries one and N two pairs of lone pairs. At the critical concentration of stabilization, the radial N:H–O tension by the circumferential H↔H repulsion not only stabilizes the *cyclo-*$N_5^-$ complexes but also stores excessive bond energy. The configuration shortens all the covalent bonds and lengthens the N:H nonbond, see context for discussion (reprinted with permission from [2, 3])

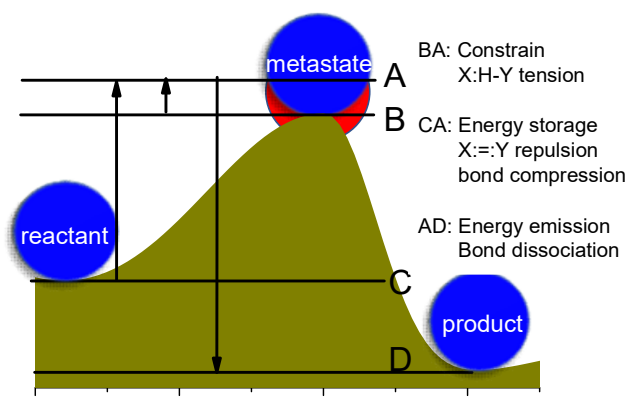

Figure 3 Energy diagram for the constrained (BA ≠ 0) and unconstrained (BA = 0) explosion. Interplay of the X:H–Y tension and the X:⇔:Y or H↔H compression stabilizes the crystal structure (BA). Intramolecular bond shortening through X:⇔:Y or H↔H compression and X:H–Y tension stores energy (CA). Tensile X:H breaking initiates and X:⇔:Y fosters detonation to eject energy (AD) whose manner and amount of energy is subject to its initial storage and structure stability.

Figure 3 summarizes the present discussion. The combination of the intermolecular X:H–Y tension and X:⇔:Y or the H↔H compression not only stabilizes the molecular structure but also stores energy by shortening the intramolecular bonds. The X:H constraints and the X:⇔:Y fosters the exploration. The system will collapse at once to erupt energy once the tensile X:H breaks. Short of X:H–Y tension endows the unconstrained explosion of alkali metals and molten alkali halides; absence of the X:⇔:Y compressions fosters no explosion of alkali metal in liquid $NH_3$. One should be focused more on the stability and energy before calculating the impact of energy eruption. As the source of life, the electron lone pairs and protons and the forms of their interactions are the key to water, ice, aqueous solutions, energetic carrier for explosion, and beyond.

Declaration
No conflicting interest is declared.

Acknowledgement
Financial support received from the National Natural Science Foundation (No21875024;11604017) of China, is gratefully acknowledged.